# Dynamical interplay between superconductivity and charge-density-wave: a nonlinear terahertz study of coherently-driven 2*H*-NbSe$_2$ and La$_{2-x}$Sr$_x$CuO$_4$

Liwen Feng[1,2,3,4†], Jiayuan Cao[1†], Tim Priessnitz[2], Yunyun Dai[5], Thales de Oliveira[6], Jiayu Yuan[7], Min-Jae Kim[2,3,4], Min Chen[6], Alexey N. Ponomaryov[6], Igor Ilyakov[6], Haotian Zhang[1], Yongbo Lv[1], Valentina Mazzotti[2,4], Gideok Kim[2], Georg Christiani[2], Gennady Logvenov[2], Dong Wu[8], Yuan Huang[5], Jan-Christoph Deinert[6], Sergey Kovalev[6], Tao Dong[7*], Nanlin Wang[7], Stefan Kaiser[2,3,4*], Hao Chu[1,2*]

[1]*Center for Ultrafast Science and Technology, School of Physics and Astronomy, Shanghai Jiao Tong University, Shanghai 200240, China*

[2]*Max Planck Institute for Solid State Research, 70569 Stuttgart, Germany*

[3]*4th Physics Institute, University of Stuttgart, 70569 Stuttgart, Germany*

[4]*Institute of Solid State and Materials Physics, Technical University Dresden, 01062 Dresden, Germany*

[5]*Advanced Research Institute of Multidisciplinary Science, Beijing Institute of Technology, Beijing 100081, China*

[6]*Helmholtz-Zentrum Dresden-Rossendorf, Bautzner Landstr. 400, 01328 Dresden, Germany*

[7]*International Center for Quantum Materials, School of Physics, Peking University, Beijing 100871, China*

[8]*Beijing academy of quantum information science, Beijing, 100193, China*

[†] These authors contributed equally.

[*] Corresponding authors:
  taodong@pku.edu.cn,
  stefan.kaiser@tu-dresden.de,
  haochusjtu@sjtu.edu.cn

**Abstract**

2*H*-NbSe2 is an archetypal system in which superconductivity and charge-density-wave (CDW) coexist and compete macroscopically with each other. In particular, this interplay also manifests in their dynamical fluctuations. As a result, the superconducting amplitude fluctuations (i.e. Higgs mode) is pushed below the quasiparticle continuum, allowing it to become a coherent excitation observable by Raman scattering. In the present study, we coherently drive the collective oscillations of the two orders and visualize their interplay in the driven states in the time domain. We find that both collective modes contribute to terahertz third harmonic generation (THG) and their THG signals interfere below $T_c$, leading to an anti-resonance of the integrated THG signal. The dynamical Ginzburg-Landau model suggests that around the anti-resonance a periodic energy transfer between the driven Higgs oscillations and the driven CDW oscillations is possible. In addition to 2*H*-NbSe2, we also studied an underdoped $La_{2-x}Sr_xCuO_4$ (x ~ 0.12) driven beyond the perturbative regime of THG. A similar interference between two sources of THG is observed below $T_c$. While there might be additional sources of THG in these experiments, our results illustrate the roles of coupled modes in the terahertz THG process and the tantalizing possibility of coherent control via such couplings.

Superconductivity and charge-density-wave (CDW) are often found as alternative ground states of metals at low temperatures. In many materials, they tend to accompany each other as close neighbors in the thermodynamic phase diagram[1-8]. This observation has prompted a sustained investigation of their relationship from the condensed matter community: for example, whether the two orders are competing or co-operative from the macroscopic point of view. Microscopically, the link between the two orders has been investigated by the SU(2) model[9] and the more specific pair-density-wave model[10] in both general and specific material context. A prototypical system which exhibits the two orders in co-existence and facilitates the discussion is 2*H*-NbSe2 ($T_c$ ~ 7 K & $T_{CDW}$ ~ 35 K). In a narrow temperature window just below $T_c$, Raman spectroscopy studies have revealed a spectral weight transfer between two low-energy collective modes as a function of temperature, indicating their hybridization[2,3,11]. While the higher-energy mode was understood to arise from the CDW amplitude fluctuations, the lower-energy mode slightly below the pair-breaking energy (2Δ) was later recognized as the amplitude mode of the superconducting order

(i.e. Higgs mode)[12-14]. The interaction between these two modes pushes the Higgs mode below the pair-breaking continuum, making it long-lived and allowing it to appear as a coherent peak in spectroscopy experiments.

Parallel to Raman scattering investigations of the Higgs mode in 2$H$-NbSe$_2$, pump probe spectroscopy is demonstrated as another versatile tool for investigating superconducting fluctuations[15-20]. Specific to this context, use of narrow-band multicycle terahertz pulses was recently made to drive superconducting fluctuations in both conventional and high-$T_c$ superconductors, which is shown to result in terahertz third harmonic generation (THG). A significant enhancement of THG below $T_c$, with unique resonance/anti-resonance behaviors was reported and interpreted in terms of terahertz-driven Higgs oscillations. While there is still debate about the fate of the Higgs mode in a conventional superconductor and its role behind terahertz THG[21-23], a further question concerning the origin of the anti-resonance/Fano interference[22] manifested by the THG response of cuprate high-$T_c$ superconductors awaits to be answered[18,19]. In the present work, we elucidate these important questions by studying the prototypical superconducting-CDW material 2$H$-NbSe$_2$. We show that both the Higgs mode and the CDW amplitude mode contribute to THG, and that their interplay indeed leads to an anti-resonance of the total THG signal.

Our measurements are performed on a 2$H$-NbSe$_2$ sample exfoliated onto a SiO$_2$ substrate[25,26], using the 0.3 THz carrier envelope phase-stable pulses emitted by TELBE, a superradiant THz source at the ELBE center for high-power radiation sources (see Supplementary Information for set-up and sample details). The field strength of the driving pulse is ~ 30 kV/cm at maximum. Before systematic measurements, we have checked that the nonlinear optical response of the sample under such driving fields stays within the cubic/perturbative regime of THG. Since both the Higgs mode and the CDW amplitude mode are Raman-active, they couple quadratically to light in the leading order. This means that the 0.3 THz pump pulse drives collective oscillations of the two orders at 0.6 THz[17,19,27]. An additional photon from the pump pulse then scatters inelastically with the driven fluctuations, giving rise to an anti-Stokes-shifted photon of 0.9 THz (THG) frequency. By measuring the transmitted terahertz field from the sample and applying the Fourier bandpass filter to the measured waveform, we can extract the THG waveform.

To extract the THG intensity, we perform Fourier transform on the raw transmission data (containing both THG and scattered linear components) and integrate the area under the 0.9 THz peak.

Figure 1 shows the temperature-dependent THG waveforms from $2H$-NbSe$_2$. Above $T_c$, THG is weak and the waveform contains artifacts from applying the Fourier bandpass filter, residual THz background, and potentially a small contribution of THG by the mobile carrier dynamics. (We observed also a finite THG above $T_{CDW}$ with a thermally-activated behavior, i.e. it increases with increasing temperature and optical pumping, which can be understood as arising from the terahertz-driven intraband/interband current from mobile carriers. This effect likely occurs in the substrate as suggested by additional measurements. Although strong at higher temperatures, it becomes negligible below the CDW transition. See Supplementary Information for details.) Nevertheless, the Fourier transform of the raw transmission results exhibits clear above-noise level THG peaks with an order parameter-like temperature dependence in the CDW phase (Fig. 2b). (See Supplementary Information for measurements on a different $2H$-NbSe$_2$ sample on Si substrate, where THG from the CDW phase is greatly enhanced.) As temperature decreases below $T_c$, THG gradually builds up in amplitude. Interestingly, around 6.2 K the waveform exhibits a noticeable dip in the middle of the envelope. This is corroborated by the splitting of the THG FFT peak at this temperature (Fig. 2a), suggesting that there are two sources of THG with different phases. In Fig. 1 we illustrate these two THG wavelets by fitting two Gaussian-enveloped sinusoidal functions with slightly different time centers ($t_1$ and $t_2$) to the data. Although for a range of $t_1$ and $t_2$ we can fit the data equally well, the resulting phase difference between the two wavelets remains largely invariant regardless of the choices of $t_1$ and $t_2$, which is particularly so around 6.2 K where the two wavelets have similar amplitudes. Therefore, in Fig. 1 we illustrate the individual wavelets using one set of $t_1$ and $t_2$ and allow the amplitude ($A_1$ and $A_2$) and phase ($\Phi_1$ and $\Phi_2$) to freely fit. With this fitting procedure, it can be seen that one THG wavelet enhances steadily below $T_c$ and dominates over the other at low temperatures.

The above interpretation can also be inferred from the temperature dependence of the integrated THG intensity as shown in Fig. 2b. While THG from the CDW phase saturates below 20 K, another prominent source of THG sets in below $T_c$ and dominates. This is consistent with recent theory results which suggest that both the CDW and Higgs fluctuations contribute to THG, and that the latter process dominates at low temperatures[28,29]. Interestingly, in the integrated THG intensity, again we observe a conspicuous dip at 6.2 K. This feature is reproduced from repeated temperature scans and also corroborates the out-of-phase interference in the time domain. We note that previous studies on cuprate high-$T_c$ superconductors reported a similar dip in THG intensity below $T_c$ despite the lack of a clear interference pattern in the time-domain waveforms[18,19]. There, the dip is also accompanied by a simultaneous jump of the THG phase (i.e. relative to the linear drive) in a direction opposite to a resonance phase jump. (In our present study we could not extract the THG phase relative to the linear drive because the terahertz transmission contains only randomly scattered linear driving field due to the use of two 0.9 THz bandpass filters after the sample for suppressing the linear leakage and improving the THG signal-to-noise ratio.) These features were recognized as an anti-resonance/Fano interference, purportedly arising from the interaction between the Higgs fluctuations and the CDW amplitude fluctuations in cuprates, although an unambiguous identification of the latter across different spectroscopic probes has yet to be agreed upon[5,6,30,31]. In the present case of 2$H$-NbSe$_2$, both the Higgs mode and the CDW amplitude mode are clearly defined by previous experimental studies and their interactions well-recognized. In light of these, we interpret the interference pattern in Fig. 1 and the dip (i.e. the anti-resonance) of the integrated THG intensity in Fig. 2b as a consequence of this interaction. It is perhaps not coincident that previous Raman scattering studies observed equal spectral weight for the two modes near 6 K[2]. This is also the temperature where we observe the most visible interference pattern, presumably as THG from both collective modes attain similar amplitudes and opposite phases at this temperature.

To gain further insights on the dynamical interaction between the two orders under a periodic drive, we employ the dynamical Ginzburg-Landau model in which we introduce two coupled order parameters[14]. The model includes the static action density:

$$S_{static} = -\alpha |\Psi|^2 + \frac{\beta}{2}|\Psi|^4 - a|\Phi|^2 + \frac{b}{2}|\Phi|^4 + \lambda |\Psi|^2 |\Phi|^2,$$

and the dynamical part

$$S_{dynamic} = -K \frac{\partial}{\partial t}\Psi^* \frac{\partial}{\partial t}\Psi - Q \frac{\partial}{\partial t}\Phi^* \frac{\partial}{\partial t}\Phi.$$

The equations of motion can be derived by introducing fluctuations in the amplitudes of the two order parameters ($\delta_\Psi$ and $\delta_\Phi$) and minimizing the action density with respect to these fluctuations. After Fourier transforming to the frequency domain, we obtain:

$$K\omega^2 \delta_\Psi + (2\alpha - 2\lambda|\Phi|^2)\delta_\Psi + 4\lambda|\Phi||\Psi|\delta_\Phi = 0,$$

$$Q\omega^2 \delta_\Phi + (2a - 2\lambda|\Psi|^2)\delta_\Phi + 4\lambda|\Phi||\Psi|\delta_\Psi = 0,$$

which can be shown as equivalent to the equations of motion of the coupled oscillators model by setting $\omega_\Psi = \sqrt{2\frac{\alpha - \lambda|\Phi|^2}{K}}$, $\omega_\Phi = \sqrt{2\frac{a - \lambda|\Psi|^2}{Q}}$ and $g_{\Psi,\Phi} = \frac{4\lambda|\Phi||\Psi|}{K}$, $g_{\Phi,\Psi} = \frac{4\lambda|\Phi||\Psi|}{Q}$:

$$(\omega^2 + \omega_\Psi^2 + i\omega\gamma_\Psi)\delta_\Psi + g_{\Psi,\Phi}\delta_\Phi = 0,$$

$$(\omega^2 + \omega_\Phi^2 + i\omega\gamma_\Phi)\delta_\Phi + g_{\Phi,\Psi}\delta_\Psi = 0.$$

Above we have also artificially introduced the damping factor $\gamma_\Psi$ and $\gamma_\Phi$ to describe the finite lifetimes of the collective excitations. One further caveat about our experiment is that the scattering cross section symmetrically involved in the 2ω driving and 3ω emission processes differ significantly between the light-Higgs and light-CDW amplitude mode scattering processes. This can be conveniently accounted for by setting the driving forces on each mode (currently set to 0 in the above equations) and introducing an identical amplification factor for each mode's oscillatory response.

Using the model described as above, we now try to reproduce the experimentally observed anti-resonance between the two driven collective modes. In particular, we make realistic choices for the temperature dependence of the parameters involved in the model (e.g. resonant frequencies of the hybridized modes, coupling constant, damping factor of the CDW amplitude mode) by referencing the Raman scattering results[2] (see Supplementary Information for details). The damping factor of the Higgs mode is set to a very small value in accordance with theoretical expectation that it is the lowest-energy excitation in the superconducting ground state of 2*H*-NbSe$_2$. The only parameters that we arbitrarily choose are the symmetric driving forces and response amplification factors, which are chosen so as to reproduce the experimentally observed Raman intensity of the two modes. With these parameters, we calculate the response of the two coupled modes under a Gaussian-enveloped periodic drive at 2ω (ω = 0.3 THz) frequency and inverse-

Fourier-transform their response into time domain. Figure 3 shows that this basic model reproduces the experimentally observed interference pattern remarkably well. Interestingly, with this model we observe a visible suppression of the THG envelope in the middle of *the CDW response* (apart from the destructive interference/suppression in the middle of *the total THG envelope*). In fact, near the anti-resonance of the heavily damped CDW amplitude mode, this kind of suppression is always present to some degree in its oscillatory response. It originates from (embodies) the very anti-resonance itself: the periodically-driven heavily damped CDW mode maximally transfers its energy to the underdamped Higgs mode, leading to a suppression of its own oscillatory response. In the meantime, as the Higgs mode reaches maximum amplitude, it feeds energy back to the CDW mode through the same coupling. This causes the latter's response to rise up again, leading to the dip in its response envelope. Under a periodic drive, such a dynamical energy transfer between the two coupled modes will always take place near the anti-resonance of the heavily damped mode. Although we cannot ascertain to what extent this effect is present in the total THG response of 2$H$-NbSe$_2$ at 6.2 K based on our current results, future experiments with more carefully tailored driving field profiles and improved signal-to-noise ratio may demonstrate this effect more clearly.

Prior to our investigation of 2$H$-NbSe$_2$, we have also studied an underdoped La$_{2-x}$Sr$_x$CuO$_4$ ($x \sim 0.12$, $T_c \sim 26$ K) under intense 0.7 THz drive. Earlier studies have shown that in the weak perturbative regime, THG from the underdoped La$_{2-x}$Sr$_x$CuO$_4$ exhibits a smooth temperature dependence. Its waveform retains a single Gaussian-like envelope until the THG signal vanishes among noise level slightly above $T_c$. When we increase the driving field strength beyond the perturbative regime of THG, the underdoped La$_{2-x}$Sr$_x$CuO$_4$ starts to exhibit interesting results similar to 2$H$-NbSe$_2$. In particular, a distinct interference pattern between two sources of THG also appears at higher driving field strengths. As Fig. 4a shows, at 25 kV/cm maximum field strength the interference becomes visible at a temperature as low as 5 K and persists up to $T_c$, beyond which a single THG source remains or dominates. The integrated THG intensity shows a monotonic decrease with increasing temperature, somewhat similar to 2$H$-NbSe$_2$. While it still stays above noise level at 150 K, we note that above 75 K the temperature dependence of THG seems to become flat. This is also within proximity of the CDW onset-temperature as previously reported by diffraction studies and spectroscopy experiments[5,7,27]. Therefore, we speculate that the two sources of THG also originate from the terahertz-driven Higgs and CDW amplitude fluctuations. However, the fact that the interference pattern becomes manifest only in the non-perturbative regime of THG in $La_{2-x}Sr_xCuO_4$ probably suggests that the superconducting fluctuations couples nonlinearly to the CDW amplitude fluctuations, whereas they couple linearly in 2$H$-NbSe$_2$ as indicated by Raman scattering results. Therefore, in $La_{2-x}Sr_xCuO_4$ the THG response of the CDW amplitude fluctuations becomes visible only when we drive the Higgs oscillations to very large amplitudes. We note that the hybridization between the two amplitude modes in cuprates has been investigated theoretically[32], and that the picture of nonlinear coupling between collective modes has also been proposed in the context of nonlinear phononics in correlated materials[33].

To conclude, we have studied the THG response of 2$H$-NbSe$_2$ under a periodic terahertz drive and observed THG contributions from both the driven CDW amplitude oscillations and Higgs oscillations. The interaction between the two collective modes leads to a Fano interference of their THG signals, which manifests as an out-of-phase interference in the time-domain. A very similar phenomenon is also manifested by the underdoped $La_{2-x}Sr_xCuO_4$ under non-perturbative drive, opening up future possibilities of investigating intertwined orders through their driven dynamics. Importantly, in contrast to conventional spectroscopy experiments, we find that in a coherent drive experiment two coupled modes may dynamically and periodically transfer energy in-between near the anti-resonance of the heavily damped mode. This opens up the tantalizing possibility of coherent control via optical pumping of coupled modes, which has been demonstrated in the scheme of nonlinear phononics but which may realize more generally among different types of collective modes of condensed phases.

The authors thank the ELBE team for the operation of the TELBE facility, Sida Tian for valuable discussions, and Ryosuke Oka for helping with sample characterization. This work is supported by National Natural Science Foundation of China (No. 12274286, No. 11888101, No. 62022089, No. 52272135), National Key Research and Development Program of China (2021YFA1400200, 2019YFA0308000), the Strategic Priority Research Program (B) of the Chinese Academy of Sciences (XDB33000000).

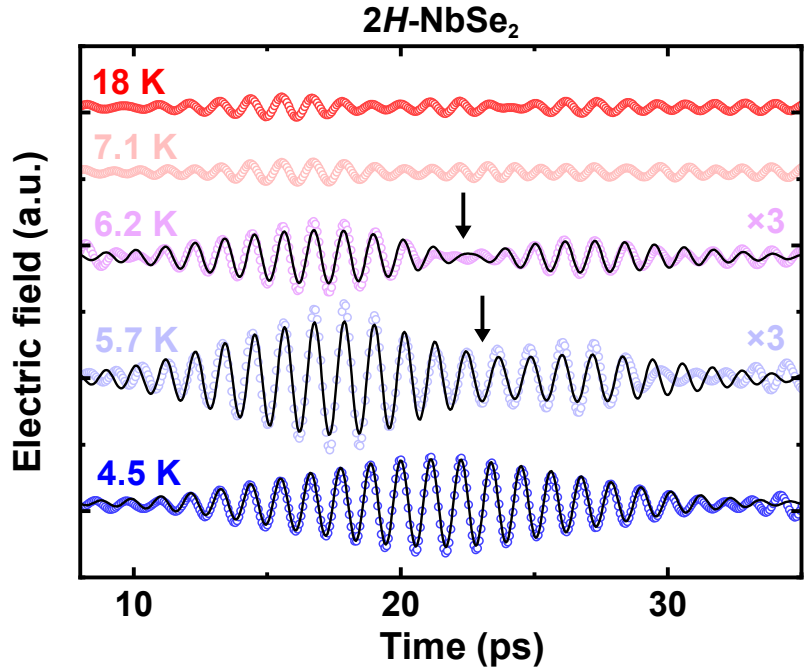

**Figure 1 Third harmonic generation (THG) from 2H-NbSe₂.** The terahertz THG experiment is performed with 0.3 THz driving frequency and placing two 0.9 THz bandpass filters after the sample for blocking the transmitted linear driving field. For extracting the THG waveforms, we apply 0.9 THz Fourier bandpass filter in our analysis. The resulting waveforms (colored circles) are shown for a few representative temperatures across $T_c$. For the three waveforms below $T_c$, we fit the data to two individual Gaussian-enveloped sinusoidal functions with fixed time centers (black lines). Near 6 K, a clear interference pattern is seen, as marked by the black arrows.

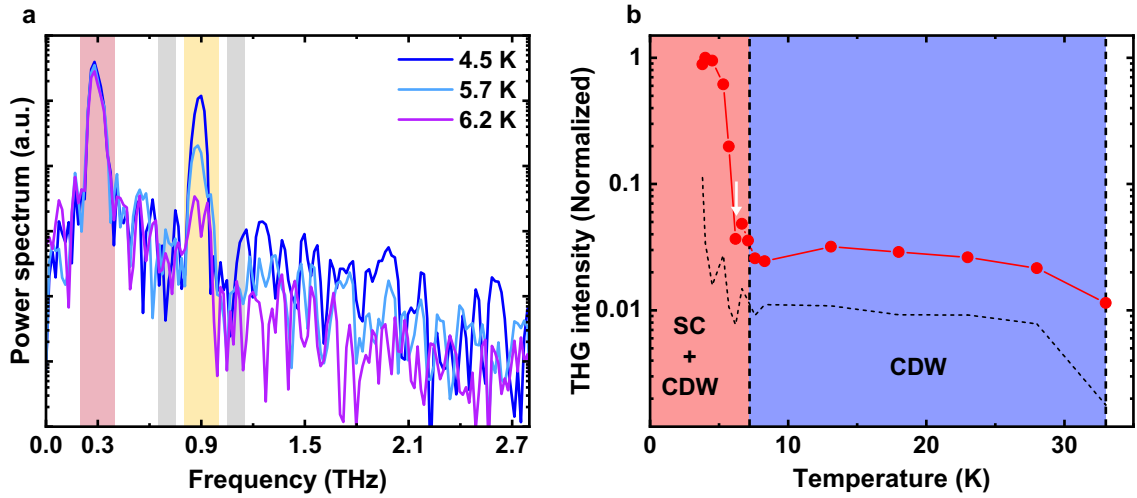

**Figure 2 Temperature dependence of THG from 2H-NbSe₂.** (a) FFT power spectra of 2H-NbSe$_2$ under a 0.3 THz multicycle drive. The maroon- and yellow-shaded areas highlight the spectra of the linear driving pulse and the THG response. The THG intensity is integrated from the yellow-shaded frequency window. The THG noise floor is estimated by integrating the grey-shaded frequency windows. Around 6.2 K a noticeable splitting of 0.9 THz peak is observed. (b) Temperature dependence of THG intensity across the superconducting transition ($T_c \sim 7$ K) up to the CDW transition ($T_{CDW} \sim 33$ K). A noticeable dip in the integrated THG intensity is manifested at 6.2 K (white arrow), in correspondence with the interference pattern in the time-domain waveform. Dotted line marks the THG noise level extracted from the FFT spectra. For a linear-scale plot of the THG intensity as a function of temperature, please see Supplementary Information.

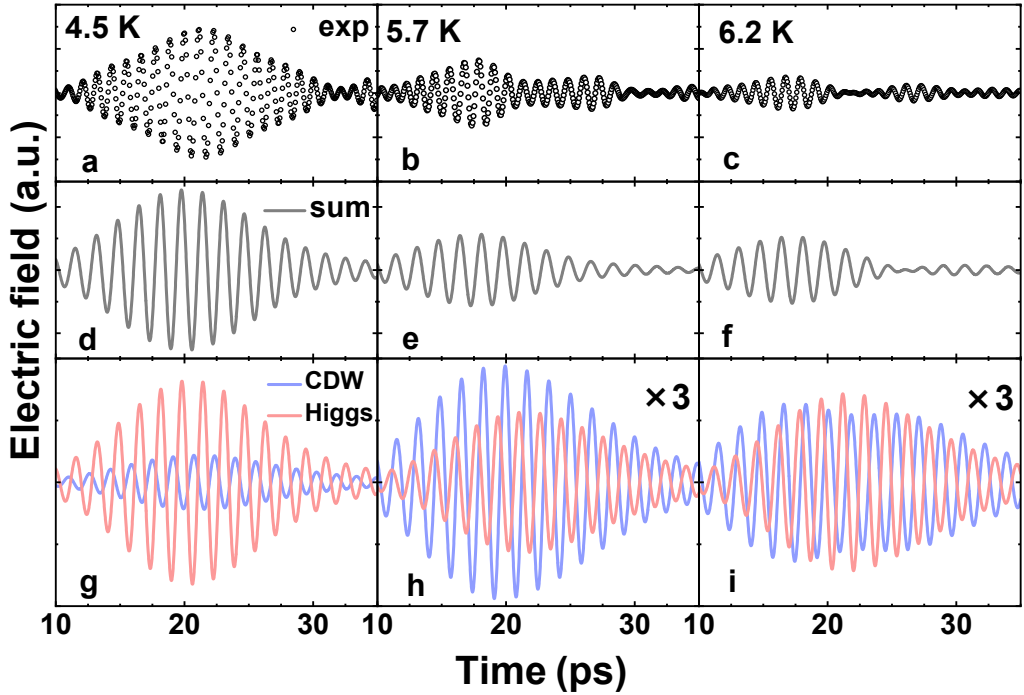

**Figure 3 Interference between two sources of THG.** **(a)-(c)** Raw THG waveforms reproduced from Fig. 1. **(d)-(f)** Coherent sums of the oscillations of two driven coupled modes as calculated from the dynamical Ginzburg Landau model described in the text. **(g)-(i)** The oscillatory response of the individual coupled modes under a Gaussian-enveloped periodic drive. In (i) the heavily damped CDW mode is undergoing an anti-resonance. A visible dip is manifested near the middle of the its response envelope (i.e. around 20 ps). It originates from the periodic energy transfer between the two coupled modes around the anti-resonance of the heavily damped CDW mode.

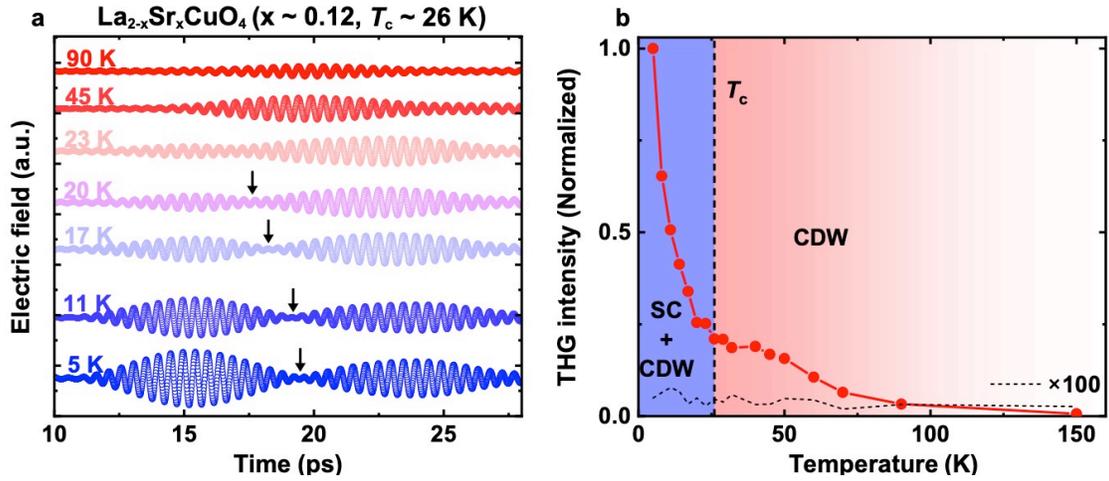

**Figure 4     Temperature dependence of THG from La$_{2-x}$Sr$_x$CuO$_4$ (x ~ 0.12).** **(a)** THG waveforms from an underdoped La$_{2-x}$Sr$_x$CuO$_4$ thin film in the non-perturbative drive regime across $T_c$ = 26 K. As marked by the black arrows, two sources of THG interfere out-of-phase with each other below $T_c$. On approaching $T_c$, one source of THG vanishes (or becomes negligible) while the other stays nonzero to much higher temperatures. **(b)** Temperature dependence of the integrated THG intensity. Dotted line marks the THG noise levels extracted from the FFT spectra.

**References**

1. A. Chikina, A. Fedorov, D. Bhoi, V. Voroshnin, E. Haubold, Y. Kushnirenko, K. H. Kim, S. Borisenko. Turning charge-density waves into Cooper pairs. *npj Quantum Materials* **5**, 22 (2020).

2. M.-A. Méasson, Y. Gallais, M. Cazayous, B. Clair, P. Rodière, L. Cario, A. Sacuto. Amplitude Higgs mode in the 2$H$-NbSe$_2$ superconductor. *Phys. Rev. B* **89**, 060503 (2014).

3. R. Grasset, T. Cea, Y. Gallais, M. Cazayous, A. Sacuto, L. Cario, L. Benfatto, M.-A. Méasson. Higgs-mode radiance and charge-density-wave order in 2$H$−NbSe$_2$. *Phys. Rev. B* **97**, 094502 (2018).

4. R. Grasset, Y. Gallais, A. Sacuto, M. Cazayous, S. Mañas-Valero, E. Coronado, M.-A. Méasson. Pressure-Induced Collapse of the Charge Density Wave and Higgs Mode Visibility in 2$H$−TaS$_2$. *Phys. Rev. Lett.* **122**, 127001 (2019).

5. J. P. Hinton, J. D. Koralek, Y. M. Lu, A. Vishwanath, J. Orenstein, D. A. Bonn, W. N. Hardy, R. Liang. New collective mode in YBa$_2$Cu$_3$O$_{6+x}$ observed by time-domain reflectometry. *Phys. Rev. B* **88**, 060508(R) (2013).

6. D. H. Torchinsky, F. Mahmood, A. T. Bollinger, I. Božović, N. Gedik. Fluctuating charge-density waves in a cuprate superconductor. *Nat. Mat.* **12**, 387–391 (2013).

7. H. Chu, L. Zhao, A. de la Torre, T. Hogan, S. D. Wilson, D. Hsieh. A charge density wave-like instability in a doped spin–orbit-assisted weak Mott insulator. *Nat. Mat.* **16**, 200-203 (2017).

8. B. Keimer, S. A. Kivelson, M. R. Norman, S. Uchida, J. Zaanen. From quantum matter to high-temperature superconductivity in copper oxides. *Nature* **518**, 179-186 (2015).

9. C. Pépin, D. Chakraborty, M. Grandadam, S. Sarkar. Fluctuations and the Higgs Mechanism in Underdoped Cuprates. *Annu. Rev. Condens. Matter Phys.* **11**, 301−323 (2020).

10. D. F. Agterberg, J.C. S. Davis, S. D. Edkins, E. Fradkin, D. J. Van Harlingen, S. A. Kivelson, P. A. Lee, L. Radzihovsky, J. M. Tranquada, Y. Wang. The Physics of Pair-Density Waves: Cuprate Superconductors and Beyond. *Annu. Rev. Condens. Matter Phys.* **11**, 231−270 (2020).

11. R. Sooryakumar & M. V. Klein. Raman Scattering by Superconducting-Gap Excitations and Their Coupling to Charge-Density Waves. *Phys. Rev. Lett.* **45**, 660 (1980)

12. P. B. Littlewood & C. M. Varma. Amplitude collective modes in superconductors and their coupling to charge-density-wave. *Phys. Rev. B* **26**, 4883−4893 (1982).

13. T. Cea & L. Benfatto. On the nature of the Higgs (amplitude) mode in the coexisting superconducting and charge-density-wave state. *Phys. Rev. B* **90**, 224515 (2014).

14. D. Pekker & C. M. Varma. Amplitude/Higgs Modes in Condensed Matter Physics. *Annu. Rev. Condens. Matter Phys.* **6**, 269−297 (2015).

15. B. Mansart, J. Lorenzana, A. Mann, A. Odeh, M. Scarongella, M. Chergui, F. Carbone. Coupling of a high-energy excitation to superconducting quasiparticles in a cuprate from coherent charge fluctuation spectroscopy. *Proc. Natl. Acad. Sci. USA* **110**, 4539−4544 (2013).

16. R. Matsunaga, Y. I. Hamada, K. Makise, Y. Uzawa, H. Terai, Z. Wang, R. Shimano. Higgs Amplitude Mode in the BCS Superconductors. *Phys. Rev. Lett.* **111**, 057002 (2013).

17. R. Matsunaga, N. Tsuji, H. Fujita, A. Sugioka, K. Makise, Y. Uzawa, H. Terai, Z. Wang, H. Aoki, R. Shimano. Light-induced collective pseudospin precession resonating with Higgs mode in a superconductor. *Science* **345**, 1145−1149 (2014).

18. H. Chu, M.-J. Kim, K. Katsumi, S. Kovalev, R. D. Dawson, L. Schwarz, N. Yoshikawa, G. Kim, D. Putzky, Z. Z. Li, H. Raffy, S. Germanskiy, J.-C. Deinert, N. Awari, I. Ilyakov, B. Green, M. Chen, M. Bawatna, G. Cristiani, G. Logvenov, Y. Gallais, A. V. Boris, B. Keimer, A. P. Schnyder, D. Manske, M. Gensch, Z. Wang, R. Shimano, S. Kaiser. Phase-resolved Higgs response in superconducting cuprates. *Nat. Commun.* **11** 1793 (2020).

19. H. Chu, S. Kovalev, Z. X. Wang, L. Schwarz, T. Dong, L. Feng, R. Haenel, M.-J. Kim, P. Shabestari, L. P. Hoang, K. Honasoge, R. D. Dawson, D. Putzky, G. Kim, M. Puviani, M. Chen, N. Awari, A. N. Ponomaryov, I. Ilyakov, M. Bluschke, F. Boschini, M. Zonno, S. Zhdanovich, M. Na, G. Christiani, G. Logvenov, D. J. Jones, A. Damascelli, M. Minola, B. Keimer, D. Manske, N. Wang, J.-C. Deinert, S. Kaiser. Fano interference of the Higgs mode in cuprate high-$T_c$ superconductor. https://arxiv.org/abs/2109.09971

20. J. Y. Yuan, L. Y. Shi, L. Yue, B. H. Li, Z. X. Wang, S. X. Xu, T. Q. Xu, Y. Wang, Z. Z. Gan, F. C. Chen, Z. F. Lin, X. Wang, K. Jin, X. B. Wang, J. L. Luo, S. J. Zhang, Q. Wu, Q. M. Liu, T. C. Hu, R. S. Li, X. Y. Zhou, D. Wu, T. Dong, N. L. Wang. Revealing strong coupling of collective modes between superconductivity and pseudogap in cuprate superconductor by terahertz third harmonic generation. https://arxiv.org/abs/2211.06961v1

21. T. Cea, C. Castellani, L. Benfatto. Nonlinear optical effects and third-harmonic generation in superconductors: Cooper pairs versus Higgs mode contribution. *Phys. Rev. B* **93**, 180507(R) (2016).

22. N. Tsuji, Y. Murakami, H. Aoki. Nonlinear light–Higgs coupling in superconductors beyond BCS: Effects of the retarded phonon-mediated interaction. Phys. Rev. B 94, 224519 (2016).

23. F. Gabriele, M. Udina, L. Benfatto. Nonlinear Terahertz driving of plasma waves in layered cuprates. *Nat. Commun.* **12**, 752 (2021).

24. M. F. Limonov, M. V. Rybin, A. N. Poddubny, Y. S. Kivshar. Fano Resonances in photonics. *Nat. Photon.* **11**, 543−554 (2017).

25. Y. Huang, E. Sutter, N. N. Shi, J. Zheng, T. Yang, D. Englund, H.-J. Gao, P. Sutter. Reliable Exfoliation of Large-Area High-Quality Flakes of Graphene and Other Two-Dimensional Materials. *ACS Nano*, **9** (11), 10612 (2015).

26. Y. Huang, Y.-H. Pan, R. Yang, L.-H. Bao, L. Meng, H.-L. Luo, Y.-Q. Cai, G.-D. Liu, W.-J. Zhao, Z. Zhou, L.-M. Wu, Z.-L. Zhu, M. Huang, L.-W. Liu, L. Liu, P. Cheng, K.-H. Wu, S.-B. Tian, C.-Z. Gu, Y.-G. Shi, Y.-F. Guo, Z. G. Cheng, J.-P. Hu, L. Zhao, G.-H. Yang, E. Sutter, P. Sutter, Y.-L. Wang, W. Ji, X.-J. Zhou & H.-J. Gao. Universal mechanical exfoliation of large-area 2D crystals. *Nat. Commun.* **11**, 2453 (2020).

27. K. Katsumi, N. Tsuji, Y. I. Hamada, R. Matsunaga, J. Schneeloch, R. D. Zhong, G. D. Gu, H. Aoki, Y. Gallais, R. Shimano. Higgs Mode in the *d*-Wave Superconductor $Bi_2Sr_2CaCu_2O_{8+x}$ Driven by an Intense Terahertz Pulse. *Phys. Rev. Lett.* **120**, 117001 (2018).

28. M. Udina, T. Cea, L. Benfatto. Theory of coherent-oscillations detection in THz pump-probe spectroscopy: from phonons to electronic collective modes. *Phys. Rev. B* **100**, 165131 (2019).

29. L. Schwarz, R. Haenel, D. Manske. Phase signatures in the third-harmonic response of Higgs and coexisting modes in superconductors. *Phys. Rev. B* **104**, 174508 (2021).

30. S. Sugai, Y. Takayanagi, N. Hayamizu. Phason and Amplitudon in the Charge-Density-Wave Phase of One-Dimensional Charge Stripes in $La_{2-x}Sr_xCuO_4$. *Phys. Rev. Lett.* **96**, 137003 (2006).

31. B. Loret, N. Auvray, Y. Gallais, M. Cazayous, A. Forget, D. Colson, M.-H. Julien, I. Paul, M. Civelli, A. Sacuto. Intimate link between charge density wave, pseudogap and superconducting energy scales in cuprates. *Nat. Phys.* **15**, 771–775 (2019).

32. Z. M. Raines, V. G. Stanev, V. M. Galitski. Hybridization of Higgs modes in a bond-density-wave state in cuprates. *Phys. Rev. B* **92**, 184511 (2015).

33. M. Först, C. Manzoni, S. Kaiser, Y. Tomioka, Y. Tokura, R. Merlin, A. Cavalleri. Nonlinear phononics as an ultrafast route to lattice control. *Nat. Phys.* **7**, 854−856 (2011).

Supplemental Information

# Dynamical interplay between superconductivity and charge-density-wave: a nonlinear terahertz study of coherently-driven 2*H*-NbSe$_2$ and La$_{2-x}$Sr$_x$CuO$_4$

Liwen Feng, Jiayuan Cao, Tim Priessnitz, Yunyun Dai, Thales de Oliveira, Jiayu Yuan, Min-Jae Kim, Min Chen, Alexey N. Ponomaryov, Igor Ilyakov, Haotian Zhang, Yongbo Lv, Valentina Mazzotti, Gideok Kim, Georg Christiani, Gennady Logvenov, Dong Wu, Yuan Huang, Jan-Christoph Deinert, Sergey Kovalev, Tao Dong, Nanlin Wang, Stefan Kaiser, Hao Chu

**S1. Experimental setup**

**S2. Sample characterization**

**S3. Temperature dependence of THG intensity from 2*H*-NbSe$_2$**

**S4. Additional contributions to THG**

**S5. Wavelets fitting procedure**

**S6. Dynamical Ginzburg-Landau model**

1. **Experimental setup**

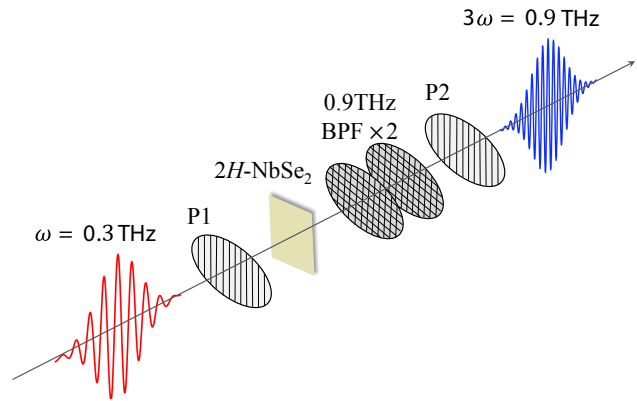

**Figure S1  Schematic of the terahertz third harmonic generation set-up.**

We performed the third harmonic generation (THG) measurements using the set-up schematically shown in Figure S1. A multicycle, phase-resolved THz source with a narrow bandwidth and high electric field strength is produced from the TELBE super-radiant undulator source at HZDR. The driving frequency is set to 0.3 THz. Two 0.9 THz bandpass filters are placed after the sample to improve the THG-to-linear signal ratio as well as the THG-to-noise ratio. Two polarizers are also placed before and after the sample for selecting parallel polarized THG for measurement. The terahertz beam is focused to a spot size on the order of 1 mm FWHM onto the sample.

For electro-optical sampling, we use a 2 mm ZnTe crystal and 100 fs gating pulses with a central wavelength of 800nm. The accelerator-based THz pulse and the 800 nm gating pulse have a timing jitter characterized by a standard deviation of ~ 20 fs. Synchronization is achieved through pulse-resolved detection.

## 2. Sample characterization

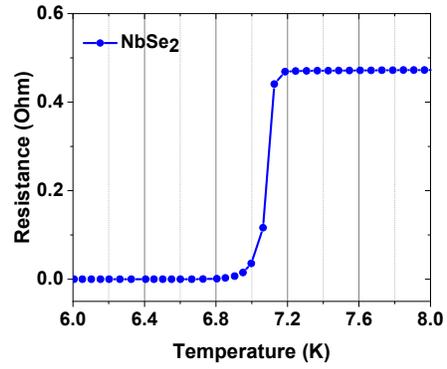

**Figure S2** **(a) 2*H*-NbSe₂ sample and (b) determination of *T*c via resistance measurement.**

The 2$H$-NbSe$_2$ sample reported in the main text of this work is exfoliated onto a SiO$_2$ substrate. The average sample thickness is about 70 nm. We performed resistance measurement on this sample and found that $T_c$ is ~ 7.2 K, indicating that the sample is pristine and in the bulk limit.

## 3. Temperature dependence of THG intensity from 2H-NbSe$_2$

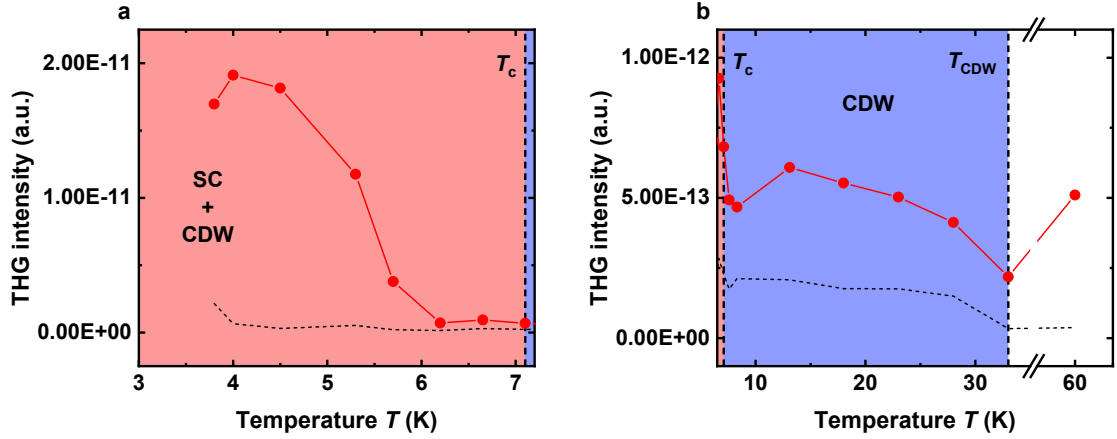

**Figure S3** Temperature dependence of THG intensity from 2H-NbSe$_2$ in **(a)** superconducting phase and **(b)** CDW phase. Black dotted line indicates the noise floor.

In 2H-NbSe$_2$, the maximum THG below $T_c$ is about two orders of magnitude stronger than THG from the CDW phase. In order to visualize the complete temperature dependence of THG below $T_{CDW}$ in a single plot, in the main text we plot the THG intensity on logarithmic scale. Here, we show the same data on linear scale for the two temperature ranges separately. From Fig. S3b, it is clear that as temperature decreases THG from the CDW phase shows an order parameter-like increase until around $T_c$, where it suddenly drops. Then THG intensity increases rapidly below $T_c$ ~ 7 K (Fig. S3a), coinciding with the superconducting transition of bulk 2H-NbSe$_2$. The THG intensity peaks at 3.7 K, which may arise from the resonance of the periodically driven Higgs mode at $2\Delta = 2\omega$, which was similarly observed in NbN superconductors.

Above $T_{CDW}$, the THG intensity rises again (Fig. S3b). We attribute this to the terahertz field-driven intraband/interband current of thermally-activated carriers from the substrate. Below $T_{CDW}$, this contribution appears to be negligible (see next section for further experimental results and discussions). We conclude that the THG signal in the CDW phase arises mostly from the terahertz-driven CDW amplitude fluctuations.

## 4. Additional contributions to THG

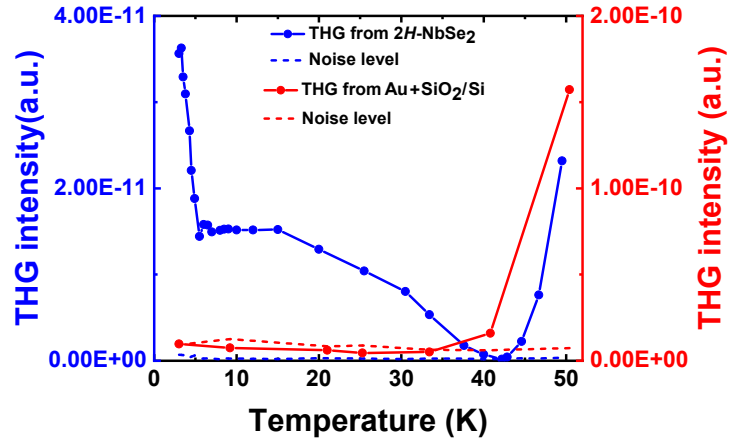

**Figure S4  Temperature dependence of THG intensity from 2*H*-NbSe$_2$ on SiO$_2$/Si substrate (blue solid line).** For comparison, we also measured THG from the bare SiO$_2$/Si substrate using a table-top terahertz source with similar field strength (red solid line). The noise floors are shown as dotted lines.

In addition to the 2*H*-NbSe$_2$ sample on SiO$_2$ substrate reported in the main text, we also measured a 2*H*-NbSe$_2$ sample on SiO$_2$/Si substrate (n-type, < 0.005 Ohm.cm, SiO$_2$ thickness 300nm, Si thickness 400±10um, 5 nm gold film deposited on top of SiO$_2$) with similar 0.3 THz multicycle pulses from TELBE. The results are shown in blue in Figure S4. While the low-temperature results resemble those in the main text, above 40 K we observe a significant increase in THG intensity with temperature. In a separate measurement performed with table-top 0.5 THz pulses (generated from the monocycle terahertz pulses from LiNbO$_3$ by the usage of two 0.5 THz bandpass filters) of comparable field strength, we found that the bare SiO$_2$/Si substrate contributes a significant source of THG above ~ 40 K (red solid line in Fig. S3). Below 35 K, THG from the SiO$_2$/Si substrate drops below noise level. This suggests that the low-temperature THG is nevertheless dominated by the Higgs and CDW amplitude mode contributions. The monotonic increase of THG above 40 K suggests that it is likely related to the thermally-activated carriers in Si driven into a nonlinear intraband/interband current, a common mechanism for high harmonic generation in semiconductors. In another optical pump THG probe experiment on the SiO$_2$/Si substrate, we observed a significant increase of THG after optical pumping (to be published by Thales de Oliveira *et al.* in a separate study), which is again consistent with the interpretation that the terahertz-driven mobile carriers contribute to THG. Here, we emphasize again that this contribution from the SiO$_2$/Si substrate is negligible below $T_{CDW}$, and in the main text by using the SiO$_2$ substrate, this contribution is further suppressed at low temperatures.

## 5. Wavelets fitting procedure

To demonstrate that the THG waveform around 6.2 K results from the interference of two THG wavelets with different phases, we fit the waveform data to the following equation:,

$$F(t) = A_1 * \exp(-(t-t_1)^2/\tau^2) * \cos(2\pi f(t-t_1) + \phi_1)$$
$$+ A_2 * \exp(-(t-t_2)^2/\tau^2) * \cos(2\pi f(t-t_2) + \phi_2).$$

Here the width of the Gaussian envelope $\tau$ is set to ~ 7ps, based on the width of the linear driving pulse used for the experiment. The frequency of the oscillation $f$ is set to 0.9 THz. The time centers of the two THG wavelets $t_1$ and $t_2$ are set to 1.5 ps apart. We can fit the data equally well for their relative delays up to ~ 2.5 ps. The delay between $t_1$ and $t_2$ is likely the result of resonance/anti-resonance. This is corroborated by the dynamical Ginzburg Landau model, where both wavelets acquire significant asymmetry in their envelope functions near the anti-resonance of the heavily damped mode (Fig. 3h,i), leading to a relative delay between their maximal oscillatory response. Here, we emphasize that the above fitting procedure is only for demonstrating the presence of two THG wavelets in the measured data. It is not for extracting the realistic response/dynamics of the individual modes, which could be more complicated as indicated by the results of the dynamical Ginzburg Landau model (Fig. 3g-i).

## 6. Dynamical Ginzburg-Landau model

As discussed in the main text, the dynamics of two coupled modes within the dynamical Ginzburg Landau model can be shown as equivalent to the coupled oscillators model. The latter has been used in our previous work[1,2] to understand the Higgs response of curpate high-$T_c$ superconductors.

For a driven coupled oscillators model, the equations of motion are given by

$$\frac{d^2}{dt^2}x_1(t) + \omega_1^2 x_1(t) + \gamma_1 \frac{d}{dt}x_1(t) + g x_2(t) = F_1(t),$$

$$\frac{d^2}{dt^2}x_2(t) + \omega_2^2 x_2(t) + \gamma_2 \frac{d}{dt}x_2(t) + g x_1(t) = F_2(t),$$

where $\omega_i$ is the natural frequency of oscillator $i$ ($i = 1, 2$), $\gamma_i$ is its damping factor, $F_i(t)$ is the time-dependent driving force on oscillator $i$, $g$ is coupling constant between the two oscillators. Assuming a periodic drive $F_i(t) = \frac{1}{2}\tilde{F}_i(\omega)e^{i\omega t} + c.c.$, we find the solution to the above equations as

$$\begin{pmatrix} \tilde{A}_1(\omega) \\ \tilde{A}_2(\omega) \end{pmatrix} = \begin{pmatrix} -\omega^2 + \omega_1^2 + i\gamma_1\omega & g \\ g & -\omega^2 + \omega_2^2 + i\gamma_2\omega \end{pmatrix}^{-1} \begin{pmatrix} \tilde{F}_1(\omega) \\ \tilde{F}_2(\omega) \end{pmatrix},$$

where $\tilde{A}_i(\omega)$ gives the amplitude of the oscillator $i$ under the driving force $\tilde{F}_i(\omega)$.

The above solution assumes a continuous driving force. To account for the Gaussian-enveloped periodic drive as used in our experiment, we re-write the driving force $F_i(t)$ as

$$F_i(t) = F_i e^{-t^2/\tau^2} \cos(\omega t),$$

and Fourier-decompose it into different frequency components $\tilde{F}_i(\omega)$:

$$F_i(t) = \int_{-\infty}^{+\infty} \tilde{F}_i(\omega) e^{i\omega t} d\omega.$$

The real-time response of oscillator $i$, $x_i(t)$, to such a Gaussian-enveloped periodic drive can be found as a linear sum of its response to the different frequency components of the driving pulse:

$$x_i(t) = \int_{-\infty}^{+\infty} \tilde{A}_i(\omega) e^{i\omega t} d\omega.$$

For the results presented in Fig. 3 of the main text, we choose $\omega_1$ and $\omega_2$ so that the eigenfrequencies of the coupled system match the energy of the Higgs mode (0.58 THz~0.65 THz) and the CDW amplitude mode (1.25 THz) as observed by Raman scattering experiments[3]. We set $\gamma_2$ (damping factor of the heavily damped oscillator) to 0.5 THz, consistent with the linewidth of the CDW amplitude mode in Raman scattering experiments. $\gamma_1$ is set to 0.01, in accordance with theoretical expectation of negligible damping of the Higgs mode[4] in 2$H$-NbSe$_2$. The dynamical Ginzburg-Landau model suggests that the coupling constant $g$ is a proportional to the product of the two order parameters' amplitudes. Therefore, it is temperature-dependent and maximum when the two order parameters reach similar amplitudes (i.e. near 6 K in 2$H$-NbSe$_2$). We take the Raman spectral weight of the Higgs mode and the CDW amplitude mode as a reference for the amplitudes of the two order parameters and compute the coupling constant accordingly. It decreases from 0.4 at 6.2K to 0.35 at 4.5K. For the driving forces, we set $F_1 = 0.2$, $F_2 = 0.9$ at 6.2 K and $F_1 = 0.2$, $F_2 = 0.3$ at 4.5 K. The driving forces reflect the light-collective mode scattering cross section (i.e. coupling vertex), which is different for the Higgs mode and the CDW collective mode. In addition, this scattering cross section is involved in both the driving process and the radiation process. To mimic this effect, we multiply the response of each oscillator by the same factor $F_1$ and $F_2$ to obtain their final oscillatory response. The central driving frequency for the above calculation is set to 0.6 THz and the width of the Gaussian envelope $\tau$ is set to 7 ps, both in accordance with experimentally used linear driving waveform.

# References

1. H. Chu, M.-J. Kim, K. Katsumi, S. Kovalev, R. D. Dawson, L. Schwarz, N. Yoshikawa, G. Kim, D. Putzky, Z. Z. Li, H. Raffy, S. Germanskiy, J.-C. Deinert, N. Awari, I. Ilyakov, B. Green, M. Chen, M. Bawatna, G. Cristiani, G. Logvenov, Y. Gallais, A. V. Boris, B. Keimer, A. P. Schnyder, D. Manske, M. Gensch, Z. Wang, R. Shimano, S. Kaiser. Phase-resolved Higgs response in superconducting cuprates. *Nat. Commun.* **11** 1793 (2020).

2. H. Chu, S. Kovalev, Z. X. Wang, L. Schwarz, T. Dong, L. Feng, R. Haenel, M.-J. Kim, P. Shabestari, L. P. Hoang, K. Honasoge, R. D. Dawson, D. Putzky, G. Kim, M. Puviani, M. Chen, N. Awari, A. N. Ponomaryov, I. Ilyakov, M. Bluschke, F. Boschini, M. Zonno, S. Zhdanovich, M. Na, G. Christiani, G. Logvenov, D. J. Jones, A. Damascelli, M. Minola, B. Keimer, D. Manske, N. Wang, J.-C. Deinert, S. Kaiser. Fano interference of the Higgs mode in cuprate high-$T_c$ superconductor. https://arxiv.org/abs/2109.09971

3. M.-A. Méasson, Y. Gallais, M. Cazayous, B. Clair, P. Rodière, L. Cario, A. Sacuto. Amplitude Higgs mode in the 2$H$-NbSe$_2$ superconductor. *Phys. Rev. B* **89**, 060503 (2014).

4. R. Grasset, T. Cea, Y. Gallais, M. Cazayous, A. Sacuto, L. Cario, L. Benfatto, M.-A. Méasson. Higgs-mode radiance and charge-density-wave order in 2$H$−NbSe$_2$. *Phys. Rev. B* **97**, 094502 (2018).